\documentclass[useAMS,usenatbib,usegraphicx]{mn2e}
\input psfig.sty
\def\spose#1{\hbox to 0pt{#1\hss}}
\def\approxlt{\mathrel{\spose{\lower 3pt\hbox{$\sim$}}
        \raise 2.0pt\hbox{$$<$$}}}
\def\approxgt{\mathrel{\spose{\lower 3pt\hbox{$\sim$}}
        \raise 2.0pt\hbox{$>$}}}

\title[Disk Outflows and the Accretion Rate Gap]{Disk Outflows and the
Accretion Rate Gap}

\author[M.C. Begelman, A. Celotti] {Mitchell C. Begelman
$^1$\thanks{E-mail: mitch@jila.colorado.edu (MB); celotti@sissa.it (AC)},
Annalisa Celotti$^{2}$\footnotemark[1]\\ $^{1}$ JILA, University of
Colorado, Boulder, CO 80309-0440, USA \\ $^{2}$ International School for
Advanced Studies (SISSA/ISAS), via Beirut 4, 34014 Trieste, Italy}

\begin{document}
\maketitle

\begin{abstract}
We argue that the observed ``accretion rate gap" --- between black
holes in radio-loud active galactic nuclei (AGN) accreting at close to
the Eddington limit and those accreting at considerably lower rates
--- can be explained in terms of the adiabatic inflow-outflow (ADIOS)
scenario for radiatively inefficient accretion. Whenever the accretion
rate falls below a threshold value (corresponding to a luminosity
$L_{\rm crit}$) that depends on the viscosity parameter, $\alpha$, the
inner region of the accretion disk --- extending from the marginally
stable orbit to $\sim 1000$ Schwarzschild radii --- is susceptible to
becoming hot and radiatively inefficient. If this happens, the disk
luminosity decreases by a factor of $\sim 100$, as most of the matter
originally destined to be swallowed is instead expelled in a wind.
According to our conjecture, accretion flows onto black holes never
radiate steadily in the range $\sim 0.01 L_{\rm crit} < L < L_{\rm
crit}$, hence the inferred accretion rate gap. We expect the gap to
exist also for black holes in X-ray binaries, where it may be
responsible for state transitions and the luminosity fluctuations
associated with X-ray nova outbursts.
\end{abstract}

\begin{keywords}
  galaxies:active --- galaxies:nuclei --- quasars:general --- radio
  continuum:galaxies --- black hole physics
\end{keywords}

\section{Introduction}

Several observational indicators point to surprisingly low accretion
luminosities in low power radio galaxies and BL Lac objects, compared
to radio-loud quasars. Evidence includes the lack of strong emission
lines, limits on nuclear photon densities, and low levels of X-ray
emission (see, e.g., Urry \& Padovani 1995 for a review; Celotti,
Fabian \& Rees 1998; O'Dowd, Urry \& Scarpa 2002; Di Matteo et
al. 2003).  The radiated output is orders of magnitude lower than that
corresponding to the (Bondi) accretion rate of the detected hot gas at
the accretion radius for a 10 per cent efficiency (Fabian \& Rees
1995, Di Matteo et al. 2003, Pellegrini et al. 2003).  Limits on the
accretion luminosity are particularly stringent for Sgr A*, where the
observational limits indicate nuclear emission corresponding to a few
$10^{-9}$ of the Eddington luminosity $L_{\rm Edd}$ of the associated
black hole (e.g., Yuan, Quataert \& Narayan 2003), and M87, where the
limit from X-rays corresponds to $\sim 10^{-7}\, L_{\rm Edd}$, and
$\sim 10^{-4}$ of the Bondi luminosity for a 10 per cent radiative
efficiency (Di Matteo et al. 2003).

Recently, Marchesini, Celotti \& Ferrarese (2004) have found evidence
for bimodal behavior of the nuclear emission, in the sense that there
appears to be a paucity of sources with bolometric accretion
luminosities in the range $10^{43}\ {\rm erg \ s}^{-1}\la L_{\rm bol}
\la 10^{45} \ {\rm erg \ s}^{-1}$.  While the low nuclear luminosities
associated with accretion in low-power radio sources (and even the
origin of the FRI/FRII dchotomy) have long been ascribed to
radiatively inefficient flow scenarios (Rees et al. 1982; Begelman
1986; Fabian \& Rees 1995; Zirbel \& Baum 1995; Reynolds et
al. 1996a,b; Ghisellini \& Celotti 2001), we concentrate here on the
actual transition to such a regime.  Motivated by the findings of
Marchesini et al. (2004), we argue in this Letter that a bimodal
luminosity distribution is a natural consequence of the adiabatic
inflow-outflow (ADIOS) scenario (Blandford \& Begelman 1999, 2004) for
radiatively inefficient accretion. After summarizing the observational
evidence in Section 2, we show how bimodality can result from ADIOS
(Section 3). We discuss the observational implications of our results
in Section 4.

\section{Observational evidence for a gap in the accretion rate}

Recent work on a sample of radio-loud sources, comprising low- (FRI,
Fanaroff \& Riley 1974) and high-power (FRII) radio galaxies and
radio-loud quasars, not only finds full agreement with the indications
of low accretion luminosities in the former systems, but also shows
the presence of a bimodal behavior in their inferred accretion rate
properties (Marchesini et al. 2004).  The accretion rate in Eddington
units, $\dot m\equiv L/\eta L_{\rm Edd}$, where $\eta = L / \dot M
c^2$ is the radiative efficiency, has been estimated from the black
hole mass, inferred from its correlation with the host bulge
B-magnitude (Merritt \& Ferrarese 2001, Gebhardt et al. 2003, H\"aring
\& Rix 2004) and from the nuclear optical luminosity measured from HST
images via a bolometric correction.

It turns out that while the black hole masses span a relatively large
range, they are clustered around $10^8-10^9$ M$_{\odot}$ and there is
no systematic difference in their values between FRI radio galaxies,
FRII radio galaxies, and radio-loud quasars.  However, the
distribution in bolometric luminosity, and (even more so) in the
corresponding mass accretion rate, appears to be bimodal, with peaks
defined around $\dot{m} \sim 10^{-3}$ for FRIs and around $\dot{m}
\sim 1$ for broad line galaxies + quasars (for $\eta = 0.1$).  The
behavior of FRII narrow line radio galaxies reflects their
spectroscopic (nuclear line) properties, in the sense that
low-excitation line galaxies behave as FRI, while high excitation FRII
galaxies likely have obscured nuclei (as already proposed, e.g., by
Laing et al. 1994; Jackson \& Wall 1999; Chiaberge, Capetti \& Celotti
2002).  If the obscuration of the high excitation radio galaxies is
taken into account, the distribution in $\dot m$ reveals a region
between the two peaks, extending about two orders of magnitude, few
$10^{-3} \le \dot{m} \le$ few $10^{-1}$, which is characterized by a
marked deficiency of sources.

We stress that the sample has been selected on the basis of the
extended radio properties and thus it is not biased by the nuclear
emission. (While the whole sample is not complete, it includes a
complete subsample of nearby [$z<0.3$] 3C radio galaxies, which also
reveals the presence of such a gap.)  We refer to Marchesini et
al. (2004) for a critical discussion of the methods, the possible role
of observational biases --- which turned out not be responsible for
the deduced bimodality --- and the findings.

As briefly discussed in the above paper, one can envisage different
processes leading to such a distribution and in particular to a
transition and a gap.  Here we focus on the possibility that such a
transition and gap arise as the flow becomes radiatively inefficient
below a certain $\dot m$ and thus becomes subject to significant mass
outflow.

\section{ADIOS and the accretion gap}

Radiatively inefficient accretion, as implied by the low nuclear
luminosities in inactive and some active galaxies, might occur
naturally at low accretion rates if the accreting gas density is low
enough to inhibit the energy coupling between protons and
electrons. Under such conditions the flow radiates very inefficiently,
remaining hot and attaining a geometrically thick configuration.

Initially, it was supposed that dissipated energy would be retained by
the accreting gas and advected into the black hole (Ichimaru 1977,
Rees et al. 1982, Narayan \& Yi 1994, Abramowicz et
al.~1995). However, attempts to model these flows dynamically revealed
an inconsistency in this argument: the transport of angular momentum
through the flow deposits so much energy in the accreting gas that it
becomes unbound at all except the innermost radii (Narayan \& Yi 1994,
1995a; Blandford \& Begelman 1999, hereafter BB99).  This suggests
that enough energy --- and presumably mass --- must be lost from the
flow to keep it bound.  BB99 proposed that this mass loss could make
the accretion rate onto the black hole much smaller than the rate at
which mass is supplied at the outer boundary. According to their
``adiabatic inflow-outflow solution" (ADIOS) scenario, the mass inflow
rate through the disk declines as $\dot m \propto r^n$ between some
outer transition radius, $r_{\rm tr}$, where the flow becomes hot, and
the marginally stable orbit, $r_{\rm ms}$, where it is captured by the
black hole. The mass flux index $n$ satisfies $0 < n <1$; values in
the upper half of this range are plausible (Blandford \& Begelman
2004).

In the ADIOS model, extremely low luminosities are possible because
the reduced accretion rate compounds the intrinsic radiative
inefficiency of the flow.  However, the luminosity cannot drop below
that produced by the thin accretion disk at $r > r_{\rm tr}$,
\begin{equation}
L_{\rm tr} \sim {\dot m (r_{\rm tr}) \over r_{\rm tr}} L_E,
\label{ltr}
\end{equation}
where $r_{\rm tr}$ is expressed in units of the Schwarzschild radius.
Note that our definition of $\dot m$ differs from that used in papers
by Narayan and collaborators by one factor of the efficiency $\eta$
(for which we adopt a value of 10 per cent) at $r_{\rm ms}$, i.e., our
values of $\dot m$ should be larger than those used by Narayan by a
factor $\eta^{-1} \sim 10$. We also assume that the inner, hot flow
does not pump a large amount of energy into the thin disk, but rather
loses most of it in the wind.  As a result, the torque across the
inner edge of the thin disk at $r_{\rm tr}$ is smaller than the
Keplerian value, and we do not expect the local dissipation rate at $r
\ga r_{\rm tr}$ to be $\sim 3$ times the local release of
gravitational binding energy (Shakura \& Sunyaev 1973).

If $n \sim 1$, the hot flow region will contribute at most a
luminosity comparable to that produced in the thin disk region. If $n$
is smaller, the amount of radiation produced near the center depends
on the rate at which electrons are heated directly (as opposed to
receiving energy from the protons), e.g., by reconnection
(Bisnovatyi-Kogan \& Lovelace 1997).  We suppose that the outer disk
luminosity gives a reasonable estimate for the total luminosity in
cases where the inner disk becomes hot.

A hot accretion flow can exist at a given radius $r$ only if the local
accretion rate, $\dot m (r)$, is smaller than some critical value,
$\dot m_{\rm crit}(r)$, that depends on the Shakura-Sunyaev viscosity
parameter, $\alpha$. For $r \la 10^3$,
\begin{equation}
\dot m_{\rm crit} \sim \dot m_{\rm crit, max} \sim 10 \alpha^2
\end{equation}
(Rees et al.~1982, Narayan \& Yi 1995b, Abramowicz et al.~1995), and
is insensitive to radius provided that the electrons and ions are
thermally coupled only through Coulomb scattering (thus allowing a
``two-temperature" flow) (Narayan 1996).  The coefficient of
$\alpha^2$ depends on details of the radiation environment, magnetic
field and flow geometry, and could be as small as a few or as large as
$\sim 50$.  At $r\ga 10^3$, $\dot m_{\rm crit}(r)$ declines with
radius at a rate that lies between $\propto r^{-1/2}$ and $\propto
r^{-3/2}$, depending on the importance of line cooling (Esin 1997).
The change in slope of $\dot m_{\rm crit}(r)$ at $r\sim 10^3$ is
associated with the fact that the virial temperature is comparable to
the electron rest mass at this radius. The radial dependence of $\dot
m_{\rm crit}(r)$ is shown schematically in Fig.~1 (analogous plots for
ADAF-like solutions, i.e., without mass loss from the flow, can be
found in Narayan \& Yi 1995b; Esin, McClintock \& Narayan 1997;
Narayan, Mahadevan \& Quataert 1998; Menou et al. 1999; Yuan \&
Narayan 2004).

Steady flows with $\dot m(r) > \dot m_{\rm crit}(r)$ must be
geometrically thin and radiative, but the converse is not necessarily
true since two solutions (thin/radiative or hot/advective) can exist
down to accretion rates $\ll \dot m_{\rm crit}(r)$. We {\it
conjecture} that, whenever two solutions exist, the hot solution is
the one chosen (see also Narayan \& Yi 1995b; Esin et al. 1997,
1998). This then implies that a small, continuous change in the mass
flow through the outer accretion disk ($r > 10^3$) can lead to an
abrupt transition in the state of the inner flow and the observed
luminosity.  If $\dot m > \dot m_{\rm crit, max}$ the radiative thin
disk extends all the way to the black hole and the luminosity is high,
but if $\dot m$ drops below $\dot m_{\rm crit, max}$ by even a small
amount, the region $r_{\rm ms}< r< 10^3$ becomes an ADIOS with a
luminosity $\ga 100$ times lower.

In the next section we show that this behavior naturally explains the
luminosity statistics of AGN.

\begin{figure}
\includegraphics[width=8cm, height=8cm]{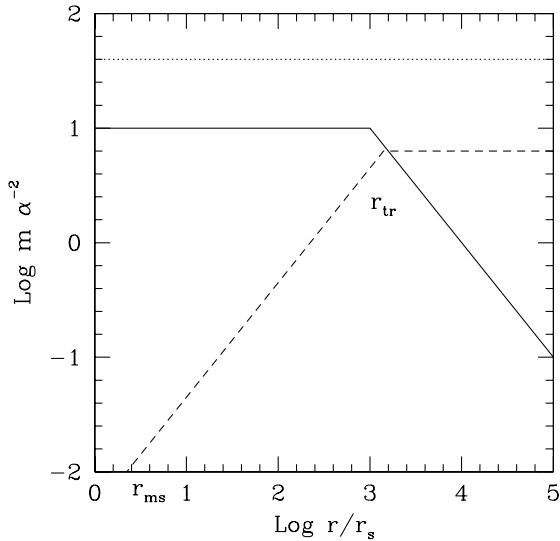}
\caption{Schematic diagram showing the accretion rate as a function of
radius, according to the ADIOS scenario. The solid curve shows the
maximum accretion rate, $\dot m_{\rm crit}$, for which a
two-temperature hot flow is possible. A flow with $\dot m > \dot
m_{\rm crit, max}$ (dotted line) exhibits a thin, radiative disk at
all radii, and conserves mass.  We conjecture that flows with $\dot m
< \dot m_{\rm crit, max}$ (dashed line) become hot within $r \ga 10^3$
and lose most of their mass before reaching the black hole.}
\end{figure}

\section{Discussion and conclusions}

If interpreted as due to ADIOS--like accretion, the observed gap in
$\dot m$ implies that for low-luminosity sources, the accretion rate
onto the black hole is about two orders of magnitude less than the
rate at the transition radius, in turn suggesting a dependence $\dot m
\propto r$. Thus, a gap in the accretion rate onto the black hole
corresponds to a continuous distribution of mass accretion rates at
the outer radii. The flow radiative inefficiency would further
contribute to -- but not solely account for -- the observed luminosity
gap.  This scenario also accounts for the discrepancy between the
(Bondi) accretion rates estimated in sources like Sgr A* and M87 and
the extremely low levels of emission detected from the nucleus.

The spectrum associated with low-luminosity accretion is expected to
be dominated by a quasi--thermal (cold) multicolor blackbody (and
lines) from $r\ge r_{\rm tr}$, with frequency peaking around a few
$10^{13}$ Hz, and a smaller total luminosity contribution from
optically thin bremsstrahlung and Compton emission plus thermal
cyclo-synchrotron radiation. The optically thin component might well
dominate at optical wavelengths, if the transition to an optically
thin outflow solution is rapid within $r_{\rm tr}$.  A non-thermal
component might also be present if a fraction of the flow energy is
dissipated into a non-thermal electron population; at present we do
not know how to calculate its magnitude. While the thermal component
is expected to be weakly variable, a non-thermal contribution produced
close to the black hole could give rise to rapidly flaring emission
(such that observed from Sgr A$^{*}$: e.g., Markoff et al 2001; Yuan
et al 2003).

A prediction specific to this scenario is the presence of ``slow"
winds associated with low $\dot m$ sources. In particular, we would
expect significant mass outflows, $\dot m_{\rm out}\sim \dot m (r_{\rm
tr}) \sim 100 \dot m(r_{\rm ms})$, with velocities $v_{\rm out} \sim
v_{\rm Kepl}(r_{\rm tr})\sim 10^4$ km s$^{-1}$, while outflows from
high--$\dot m$ sources would have speeds $\sim c$ and would emanate
from close to the black hole.

It should also be noted that the available flow energy reaching $\sim
r_{\rm ms}$ amounts typically to $\sim 10^{42-43}$ erg s$^{-1}$ for
black hole masses in the range $\sim 10^8-10^9 M_\odot$. This appears
to be barely sufficient to energize the jets associated with these
systems (e.g., M87, Owen, Eilek \& Kassim 2000, Reynolds et al. 1996a;
3C 84, Fabian et al. 2002) unless the jet forms on larger disk scales
--- as might be indicated by the high resolution imaging of M87
(Biretta, Junor \& Livio 2002) --- or involves energy extraction from
the spin of the black hole.

Finally, we note that the transition between luminosity states and the
presence of a gap in the accretion properties of radio-loud AGN could
be analogous to spectral state transitions in black hole X-ray binary
systems and microquasars.  The values of the transitional $\dot m$,
the nuclear spectral properties and possibly even the onset and nature
of jet production (e.g., Fender et al. 1999, Meier 2001, Fender 2003)
could all be broadly similar. In the X-ray binary case, in fact, it
appears that jets are associated only with accretion states (very high
and low/hard) corresponding to high and low accretion rates.  There
appear to be some differences, however.  First, the results by
Marchesini et al. (2004) show that in the case of AGN, the gap in
accretion rates is larger than that in the case of stellar-mass black
holes (where it typically spans a factor $\sim 10$, e.g., Nowak 1995,
Fender 2003). Secondly, a complete analogy between X-ray binaries and
AGN would require the identification of a population of non-jetted AGN
with intermediate accretion rates (corresponding to the high/soft
state in binaries). So far, the existence of an accretion rate gap has
been established only for radio-loud AGN.  It remains to be seen
whether intermediate accretion rates occur in the more numerous radio
quiet AGN, or if such intermediate states are entirely absent for
supermassive black holes.

\section*{Acknowledgments}

The project was partially supported by NSF grant AST-0307502 (MB) and
the Italian MIUR and INAF (AC). AC thanks the Fellows of JILA,
University of Colorado, for their warm hospitality.


\begin{thebibliography}{}

\bibitem[Abramowicz et al. (1995)]{abr95} Abramowicz M. A., Chen X.,
  Kato S., Lasota J. P., Regev O., 1995, ApJ, 438, L37

\bibitem[Begelman 1986]{MCB86} Begelman M.C., 1986, Nature, 322, 614

\bibitem[Biretta etal]{biretta} Biretta, J. A.; Junor, W.; Livio, M.,
2002, New Astr. Rev, 46, 239

\bibitem[Bisnovatyi-Kogan \& Lovelace (1997)]{BiLo97} Bisnovatyi-Kogan
  G. S., Lovelace R. V. E., 1997, ApJ, 486, L43

\bibitem[Blandford \& Begelman (1999)]{bla99} Blandford R. D.,
  Begelman M. C., 1999, MNRAS, 303, L1 (BB99)

\bibitem[Blandford \& Begelman (2004)]{bla03} Blandford R. D.,
  Begelman M. C., 2004, MNRAS, 349, 68 

\bibitem[Celotti, Fabian \& Rees 1998]{celotti98} Celotti A., Fabian
   A.C., Rees M.J., 1998, MNRAS, 293, 239

\bibitem[Chiaberge, Capetti \& Celotti 2002]{ccc02} Chiaberge M.,
  Capetti A., Celotti A., 2002, A\&A, 394, 791

\bibitem[Di Matteo et al. 2003]{dimatteo03} Di Matteo T., Allen S.W.,
  Fabian A.C., Wilson A.S., Young A.J., 2003, ApJ, 582, 133

\bibitem[Esin (1997)]{es97} Esin A.A., 1997, ApJ, 482, 400

\bibitem[Esin (1997)]{es97} Esin A.A., McClintock J.E., Narayan R.,
1997, ApJ, 489, 865

\bibitem[Esin (1997)]{es97} Esin A.A., Narayan R., Cui W., Grove J.E.,
Zhang S.-N., 1998, ApJ, 505, 854

\bibitem[Fabian \& Rees 1995]{fabian95} Fabian A.C., Rees M.J., 1995,
  MNRAS, 277, L55

\bibitem[Fabianetal]{fabianea} Fabian A.C., Celotti A., Blundell K.M.,
Kassim N.E., Perley R.A., 2002, MNRAS, 331, 369

\bibitem[Fanaroff \& Riley 1974]{fanarof74} Fanaroff B.L., Riley J.M.,
  1974, MNRAS, 167, 31

\bibitem[Fenderea]{fenderea} Fender R.P. et al. 1999, ApJ, 519, L165

\bibitem[Fender 2003]{fender} Fender R., 2003, in Compact Stellar
  X-Ray Sources, eds. W.H.G. Lewin and M. van der Klis, Cambridge
  University Press, in press (astro-ph/0303339) 

\bibitem[Gebhardt et al. 2003]{gebhardt03} Gebhardt K. et al., 2003,
  ApJ, 583, 92

\bibitem[Ghisellini G., Celotti A. 2001]{ggac01} Ghisellini G.,
Celotti A., 2001, MNRAS, 327, 739

\bibitem[H\"aring \& Rix (2004)]{haring04} H\"aring N., Rix H.-W.,
2004, ApJ, 604, L89

\bibitem[Ichimaru (1977)]{ich77} Ichimaru S., 1977, ApJ, 214, 840

\bibitem[Jackson \& Wall 1999]{jackson99} Jackson C.A., Wall J.V.,
  1999, MNRAS, 304, 160

\bibitem[Laing et al. 1994]{laing94} Laing R.A., Jenkins C.R., Wall
  J.V., Unger S.W., 1994, in The First Stromlo Symposium: The Physics
  of Active Galaxies. ASP Conference Series, Vol. 54, eds. G.V. Bicknell,
  M.A. Dopita, and P.J. Quinn, 201
 
\bibitem[Markoff S., Falcke H., Yuan F., Biermann P.L. (2001)]{mark01} 
Markoff S., Falcke H., Yuan F., Biermann P.L., 2001, A\&A, 379, L13

\bibitem[Marchesini, Celotti \& Ferrarese (2003)]{mar03} Marchesini
   D., Celotti A., Ferrarese L., 2004, MNRAS, in press (astro-ph/0403272)

\bibitem[Meier 2001]{meier01} Meier D.L., 2001, ApJ, 548, L9

\bibitem[]{} Menou K., Esin A.A., Narayan R., Garcia M.R., Lasota
 J.-P., McClintock J.E., 1999, ApJ, 520, 276
 
\bibitem[Merritt \& Ferrarese 2001]{merritt01} Merritt D., Ferrarese
  L., 2001, in The Central Kiloparsec of Starbursts and AGN: The La
  Palma Connection, ASP Conf. Series, Vol. 249, eds. J.H. Knapen,
  J.E. Beckman, I. Shlosman and T.J. Mahoney, 335

\bibitem[]{}Narayan R., Mahadevan R., Quataert E., 1998, Theory of
Black Hole Accretion Disks, eds. M.A. Abramowicz, G. Bjornsson,
J.E. Pringle, Cambridge University Press, 148

\bibitem[Narayan (1996)]{na96} Narayan R., 1996, ApJ, 462, 136

\bibitem[Narayan \& Yi (1994)]{nar94} Narayan R., Yi I., 1994, ApJ,
  428, L13

\bibitem[Narayan \& Yi (1995a)]{nar95} Narayan R., Yi I., 1995a, ApJ,
  444, 231

\bibitem[Narayan \& Yi (1995b)]{nayi95} Narayan R., Yi I., 1995b, ApJ,
  452, 710

\bibitem[novak]{novak} Novak M.A., 1995, PASP, 107, 1207

\bibitem[O'Dowd, Urry \& Scarpa 2002]{odowd02} O'Dowd M., Urry C.M.,
  Scarpa R., 2002, ApJ, 580, 960

\bibitem[Owenetal] {owen} Owen F.N., Eilek J.A., Kassim N.E., 2000,
ApJ, 543, 611

\bibitem[Pellegrini et al. 2003]{pellegrini03} Pellegrini S., Venturi
  T., Comastri A., Fabbiano G., Fiore F., Vignali C., Morganti R.,
  Trinchieri G., 2003, ApJ, 585, 677

\bibitem[Rees et al. 1982]{ree82} Rees M. J., Begelman M. C.,
  Blandford R. D., Phinney E. S., 1982, Nature, 295, 17

\bibitem[Reynolds ea]{reynolds} Reynolds C.S., Fabian A.C., Celotti
  A., Rees M.J., 1996a, MNRAS, 283, 873

\bibitem[Reynolds ea]{reynolds} Reynolds C.S., di Matteo T., Fabian
 A.C., Hwang U., Canizares C.R., 1996b, MNRAS, 283, L111

\bibitem[Shakura \& Sunyaev] {shak} Shakura N. I., Sunyaev R. A.,
1973, A\&A, 24, 337

\bibitem[Urry \& Padovani 1995]{urry95} Urry C.M., Padovani P., 1995,
  PASP, 107, 803

\bibitem[yaun]{yuan} Yuan F., Quataert E., Narayan R., 2003, ApJ, 598, 301

\bibitem[]{} Yuan F., Narayan R., 2004, ApJ, in press (astro-ph/0401117)

\bibitem[Zirbel ea]{zirbel}  Zirbel E.L., Baum S.A., 1995, ApJ, 448, 521


\end{thebibliography}
\end{document}